\documentclass[aip,apl,reprint,twocolumn,superscriptaddress,showpacs,floatfix]{revtex4-1}
\usepackage{graphicx,subfigure,amsmath,pstricks}
\usepackage[]{natbib}

\newcommand{\code}[1]{\texttt{#1}}
\newcommand{\rnd}[1]{\left(  #1 \right)}
\newcommand{\sqr}[1]{\left[  #1 \right]}

\newcommand{\TC}{T_{\mathrm{c}}}

\newcommand{\EP}{E_{\mathrm{Peierls}}}
\newcommand{\ES}{E_{\mathrm{saddle}}}
\newcommand{\EM}{E_{\mathrm{minimum}}}

\newcommand{\SP}{\sigma_{\mathrm{Peierls}}}
\newcommand{\SW}{\sigma_{\mathrm{Wall}}}

\newcommand{\unit}[1]{\,\mathrm{#1}}
\begin{document}
\pagestyle{plain}

\title{Intrinsic activation energy for twin wall motion}
\thanks{W. T. Lee, E. K. H. Salje, L. Goncalves-Ferreria,
  M. Daraktchiev and U. Bismayer. Physical Review B 73:214110
  (2006). ``Copyright (2006) by the American Physical Society.''}
\author{W.~T.~Lee}
\email{wlee@esc.cam.ac.uk}
\affiliation{Department of Earth Sciences, University of Cambridge, Downing
  Street, Cambridge, CB2 3EQ, UK.} 
\affiliation{Mineralogisch-Petrographisches Institut, Universit\"at Hamburg,
  Grindelallee 48, D-20146 Hamburg.} 
\author{E.~K.~H.~Salje}
\affiliation{Department of Earth Sciences, University of Cambridge, Downing
  Street, Cambridge, CB2 3EQ, UK.} 
\author{L.~Goncalves-Ferreira}
\affiliation{Department of Earth Sciences, University of Cambridge, Downing
  Street, Cambridge, CB2 3EQ, UK.} 
\author{M.~Daraktchiev}
\affiliation{Department of Earth Sciences, University of Cambridge, Downing
  Street, Cambridge, CB2 3EQ, UK.} 

\author{U.~Bismayer}
\affiliation{Mineralogisch-Petrographisches Institut, Universit\"at Hamburg,
  Grindelallee 48, D-20146 Hamburg.} 

\begin{abstract}
Even in a topologically perfect crystal, a moving twin wall will experience
forces 
due to the discrete nature of the lattice. The potential energy landscape can
be described in terms of one of two parameters: the Peierls energy,
which is the activation energy for domain wall motion in a perfect crystal; and
the Peierls stress, the maximum pinning stress that the potential can
exert. We investigate these parameters in a one order parameter discrete
Landau-Ginzburg model and a classical potential model of the ferroelastic
perovskite CaTiO$_3$. Using the one order parameter model we show that the
Peierls energy scales with the barrier height of the Landau double well
potential and 
calculate its dependence on the width of the wall numerically. In CaTiO$_3$
we calculate the Peierls energy and stress indirectly from the one order
parameter model and directly from the interatomic force field.
Despite the simplicity of the one order parameter model, its predictions of the
activation energy are in good agreement with calculated values.

\end{abstract}


\maketitle

\section{Introduction}

The motion of ferroelastic or ferroelectric-ferroelastic twin walls plays a
significant role in determining the elastic, dielectric, piezoelectric and
ferroelectric properties of a number of materials of scientific and
technological interest.\cite{Harrison2002, Harrison2004a, Harrison2004b,
  Moulson2003, Damjanovic1997, Scott2000} The lower mantle of the Earth is
known to consist 
mainly of magnesium silicate perovskite, a ferroelastic polymorph of
MgSiO$_3$. Recent 
work has explored the possibility that the seismic properties of the lower
mantle, such as attenuation, can be explained in terms of the elastic response
of domain 
walls. The large piezoelectric and dielectric coefficients of barium 
titanate and lead titanate have been shown to have significant contributions
from the motion of twin walls. Finally, ferroelectric switching, which is
currently being exploited for computer random access memory applications, is
known to be entirely due to the motion of twin walls.  

To understand the properties of these materials, and the systems
in  which they are found, we must understand the factors which affect the
motion of twin walls. Unlike magnetic domain walls, with widths of 100's
of nanometres, ferroelastic and ferroelectric walls are atomistically thin,
with wall widths of the order of the unit cell parameter.\cite{Salje2005a} We
must understand their behaviour from an atomic perspective. In essence this
requires an understanding of the energy landscape through which twin walls
move.  

\begin{figure}[!h]
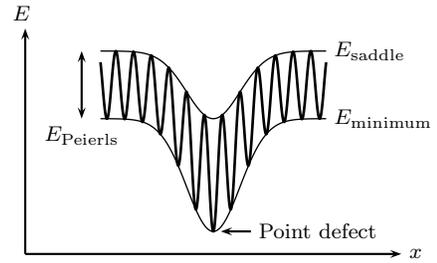

\begin{center}

\end{center}
\caption{\label{energylandscape} The energy landscape experienced by a domain
  wall moving through a crystal containing a point defect. The bold line shows
  the potential experienced 
  by the wall while the thin lines show the local equilibrium energy of the
  wall $\EM$ and the saddle point energy of the transition state from one
  local minimum to another $\ES$. Far from a defect the difference between
  these two energies, i.e. the activation energy needed for the wall to move
  is called the Peierls energy $\EP$.}
\end{figure}

A schematic of this energy landscape is shown
in Fig.~\ref{energylandscape}. As a wall moves through the crystal it
experiences a potential which oscillates between $\EM$ at local
minima and $\ES$ at transition states (saddle points) between two minima. If
the motion of the wall is a thermally activated process then the mobility of
the wall depends on the difference between these two energies.
\begin{equation}
\mu_{\mathrm{wall}}=\mu_0 \exp\rnd{-\dfrac{\ES-\EM}{kT}}
\end{equation}

 As shown in the figure there are
two contributions to the potential energy landscape. These can be labelled
\emph{intrinsic} and \emph{extrinsic}. The intrinsic 
contribution to the energy landscape is present even in a chemically and
topologically perfect crystal and is due to the periodicity of the
lattice. This contribution is parameterised by the Peierls energy $\EP$, which
is the activation energy for twin wall motion far from point defects. The
second, extrinsic contribution to the energy landscape is due to defects in the
perfect lattice such as vacancies, impurity atoms, dislocations and other twin
walls. 

There have been attempts to understand this energy landscape using both
experimental and simulation methods. Experimentally, it is clear that
extrinsic pinning due to point defects is far more significant than lattice
pinning,\cite{Harrison2004a,Harrison2004b, Scott2000}  which is often too
small to detect, except by very sensitive methods.\cite{Ma2003} Simulations
of oxygen vacancies in ferroelastic calcium titanate\cite{Calleja2003} and
ferroelectric lead zirconate\cite{He2003} have already been carried out,
showing that an oxygen vacancy has an energy approximately
$1\unit{eV}$ lower in the wall than in the bulk. These simulations
provide information about $\EM$, but in order to complete the picture,
calculations of $\ES$ are also necessary.   

In this work we only consider intrinsic pinning, We investigate pinning in a
one order parameter discrete Landau-Ginzburg model, building on previous work
by Ishibashi\cite{Ishibashi1979} and Combs and Yip\cite{Combs1983,
Combs1984}. Then we investigate intrinsic pinning in an empirical potential
model of orthorhombic calcium titanate CaTiO$_3$ developed by Calleja et
al.\cite{Calleja2003} We compare the predictions of the one order parameter
model with the results of a transition state calculation and show a good
agreement between the two values. 

\section{A one parameter model}
\label{GeneralModel}

In this section we investigate the intrinsic pinning of domain walls in the
discrete Landau-Ginzburg or $\phi^4$ model.\cite{Salje2005a} Intrinsic pinning
in this model has been investigated by Ishibashi and
Combs and Yip.\cite{Ishibashi1979,Combs1983,Combs1984}  We report the results
of a numerical calculation of the pinning
energy showing, in agreement with previous work, that when the wall
width is two lattice spacings the activation energy is practically zero. 

The most successful theoretical tool for describing phase transitions in
ferroelectric and ferroelastic materials is Landau-Ginzburg
theory.\cite{Landau80,Salje1990} Through the Landau-Ginzburg free energy 
the theory provides a framework which can be used to 
predict both macroscopic behaviour, such as the specific heat
capacity~\cite{Salje1988b} and elastic 
constants~\cite{Carpenter1998b} of a material going through a phase
transition, and microstructural details, such as the structure of domain
walls.\cite{Salje1990, Lee2003a}  

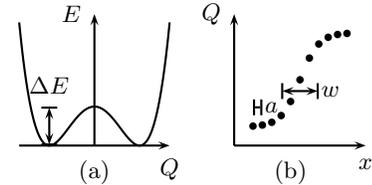
\begin{figure}[!h]
\begin{center}
 \psset{unit=0.5cm}
 \begin{pspicture}(5,5)
 \psline[linecolor=black]{->}(2.5,1.0)(2.5,4.5)
 \psline[linecolor=black]{->}(0.5,1.0)(4.5,1.0)
 \uput[180](2.5,4.5){$E$}
 \uput[270](4.5,1.0){$Q$}
 \pscurve[linecolor=black](  0.5,  4.27679971)(  0.54,  3.88384101)(  0.58,
  3.52315622)(  0.62,  3.19325544)(  0.66,  2.89267946)(  0.7,  2.61999979)(
  0.74,  2.37381869)(  0.78,  2.15276911)(  0.82,  1.95551472)(  0.86,
  1.78074994)(  0.9,  1.62719987)(  0.94,  1.49362037)(  0.98,  1.37879798)(
  1.02,  1.28154999)(  1.06,  1.20072441)(  1.1,  1.13519994)(  1.14,
  1.08388603)(  1.18,  1.04572285)(  1.22,  1.01968126)(  1.26,  1.00476287)(
  1.3,  1.)(  1.34,  1.00445569)(  1.38,  1.0172237)(  1.42,  1.03742851)(
  1.46,  1.06422532)(  1.5,  1.09680005)(  1.54,  1.13436934)(  1.58,
  1.17618055)(  1.62,  1.22151176)(  1.66,  1.26967176)(  1.7,  1.32000009)(
  1.74,  1.37186698)(  1.78,  1.42467339)(  1.82,  1.47785099)(  1.86,
  1.5308622)(  1.9,  1.58320012)(  1.94,  1.63438861)(  1.98,  1.68398221)(
  2.02,  1.73156622)(  2.06,  1.77675662)(  2.1,  1.81920015)(  2.14,
  1.85857423)(  2.18,  1.89458703)(  2.22,  1.92697744)(  2.26,  1.95551504)(
  2.3,  1.98000016)(  2.34,  2.00026384)(  2.38,  2.01616784)(  2.42,
  2.02760464)(  2.46,  2.03449744)(  2.5,  2.03680016)(  2.54,  2.03449744)(
  2.58,  2.02760464)(  2.62,  2.01616784)(  2.66,  2.00026384)(  2.7,
  1.98000016)(  2.74,  1.95551504)(  2.78,  1.92697744)(  2.82,  1.89458703)(
  2.86,  1.85857423)(  2.9,  1.81920015)(  2.94,  1.77675662)(  2.98,
  1.73156622)(  3.02,  1.68398221)(  3.06,  1.63438861)(  3.1,  1.58320012)(
  3.14,  1.5308622)(  3.18,  1.47785099)(  3.22,  1.42467339)(  3.26,
  1.37186698)(  3.3,  1.32000009)(  3.34,  1.26967176)(  3.38,  1.22151176)(
  3.42,  1.17618055)(  3.46,  1.13436934)(  3.5,  1.09680005)(  3.54,
  1.06422532)(  3.58,  1.03742851)(  3.62,  1.0172237)(  3.66,  1.00445569)(
  3.7,  1.)(  3.74,  1.00476287)(  3.78,  1.01968126)(  3.82,  1.04572285)(
  3.86,  1.08388603)(  3.9,  1.13519994)(  3.94,  1.20072441)(  3.98,
  1.28154999)(  4.02,  1.37879798)(  4.06,  1.49362037)(  4.1,  1.62719987)(
  4.14,  1.78074994)(  4.18,  1.95551472)(  4.22,  2.15276911)(  4.26,
  2.37381869)(  4.3,  2.61999979)(  4.34,  2.89267946)(  4.38,  3.19325544)(
  4.42,  3.52315622)(  4.46,  3.88384101)(  4.5,  4.27679971)
 \psline[linecolor=black]{|<->|}(  1.29999995,  2.03680016)(  1.29999995,  1.)
 \uput[90](  1.29999995,  2.03680016){\black $\Delta E$}
 \rput(2.5,0.3){(a)}
 \end{pspicture}
 \begin{pspicture}(5,5)
 \psline[linecolor=black]{->}(1.0,1.0)(1.0,4.5)
 \psline[linecolor=black]{->}(1.0,1.0)(4.5,1.0)
 \uput[180](1.0,4.5){$Q$}
 \uput[270](4.5,1.0){$x$}
 \pscircle*[linecolor=black](  1.5,  1.51673213){0.1}
 \pscircle*[linecolor=black](  1.75,  1.54496552){0.1}
 \pscircle*[linecolor=black](  2.,  1.61856468){0.1}
 \pscircle*[linecolor=black](  2.25,  1.79800731){0.1}
 \pscircle*[linecolor=black](  2.5,  2.17235355){0.1}
 \pscircle*[linecolor=black](  2.75,  2.75){0.1}
 \pscircle*[linecolor=black](  3.,  3.32764645){0.1}
 \pscircle*[linecolor=black](  3.25,  3.70199269){0.1}
 \pscircle*[linecolor=black](  3.5,  3.88143532){0.1}
 \pscircle*[linecolor=black](  3.75,  3.95503448){0.1}
 \pscircle*[linecolor=black](  4.,  3.98326787){0.1}
 \psline[linecolor=black]{|-|}(  1.5,  2.01673213)(  1.75,  2.01673213)
 \uput{1pt}[0](  1.75,  2.01673213){\black $a$}
 \psline[linecolor=black]{|<->|}(  2.25,  2.45)(  3.25,  2.45)
 \uput{1pt}[0](  3.25,  2.45){\black $w$}
 \rput(2.5,0.3){(b)}
 \end{pspicture}
 \psset{unit=1.0cm}
\end{center}
\caption{\label{doublewell} Dimensional quantities of the general model. (a)~The Landau 
  double well potential is characterised by an energy barrier
  $\Delta E$. (b)~The discrete nature of the lattice means the model contains
  two lengthscales: $a$~the lattice parameter and $w$~the wall width.}
\end{figure}

Usually a continuum formulation of the Landau-Ginzburg free energy is used, in
which the discrete nature of the lattice is
neglected.\cite{Salje1990,Salje2005a} This approach has been very successful
even in predicting the structure of twin walls, where the continuum
approximation might be expected to break down. It is not possible to calculate
the  Peierls energy within the continuum limit and so we use a discrete
form of the Landau-Ginzburg energy,    
\begin{equation}
F=\sum_i \Delta E\sqr{ \rnd{Q_i^2-1}^2 + \rnd{\dfrac{w}{a}}^2 \rnd{Q_{i+1}-Q_i}^2}
\end{equation}
The first term is a double well potential, where $\Delta E$ is height of the
barrier between the two walls. The second term is the discrete analogue of the
Ginzburg term. $w$ is the wall width and $a$ is the lattice spacing. These
quantities are illustrated in Fig.~\ref{doublewell}.

Dimensional analysis tells us that the Peierls energy $\EP$ must be given by
\begin{equation}  
\label{generalequation}
\EP=\Delta E f\rnd{w/a}
\end{equation}
where $f$ is to be determined. It is easy to deduce the limiting values of
$f(x)$ in the cases when $x$ is very small or very large. 

In
the case $w=0$ the free energy of the discrete Landau-Ginzburg model is
\begin{equation} 
F=\sum_i \Delta E \rnd{Q_i^2-1}^2
\end{equation}
The system consists of a collection of independent order parameters $Q_i$
moving in double well potentials. The domain wall moves when one order
parameter flips from one state to another. The activation energy for this
process is $\Delta E$ and thus
\begin{equation}
\lim_{x \rightarrow 0} f\rnd{x}=1
\end{equation}
If $w$ is very large then the discrete nature of the lattice is irrelevant and 
a continuum approximation may be used. In the continuum theory the energy of a
domain wall is independent of its position, and thus there is no activation
energy for domain wall motion
\begin{equation} 
\lim_{x \rightarrow \infty} f\rnd{x}=0
\end{equation}

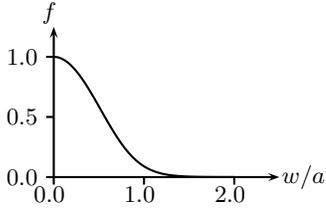
\begin{figure}[!h]
\begin{center}
 \begin{pspicture}(5,3)
\psline[linecolor=black]{->}(  1.0000,  0.5000)(  4.0000,  0.5000)
\psline[linecolor=black]{->}(  1.0000,  0.5000)(  1.0000,  2.5000)
\uput{1pt}[0](          4.000000,          0.500000){$w/a$                                   }
\uput{1pt}[90](          1.000000,          2.500000){$f$                                     }
\psline[linecolor=black](  1.0000,  0.4000)(  1.0000,  0.5000)
\psline[linecolor=black](  2.2000,  0.4000)(  2.2000,  0.5000)
\psline[linecolor=black](  3.4000,  0.4000)(  3.4000,  0.5000)
\psline[linecolor=black](  0.9000,  0.5000)(  1.0000,  0.5000)
\psline[linecolor=black](  0.9000,  1.3000)(  1.0000,  1.3000)
\psline[linecolor=black](  0.9000,  2.1000)(  1.0000,  2.1000)
\uput{1pt}[-90](          1.000000,          0.400000){\black $0.0$                                   }
\uput{1pt}[-90](          2.200000,          0.400000){\black $1.0$                                   }
\uput{1pt}[-90](          3.400000,          0.400000){\black $2.0$                                   }
\uput{0pt}[180](          0.900000,          0.500000){\black $0.0$                                   }
\uput{0pt}[180](          0.900000,          1.300000){\black $0.5$                                   }
\uput{0pt}[180](          0.900000,          2.100000){\black $1.0$                                   }
 \pscurve[linecolor=black](  1.006,  2.09992)(  1.012,  2.09968)(  1.018,
  2.09928)(  1.024,  2.09872)(  1.03,  2.098)(  1.036,  2.0971232)(  1.042,
  2.0960832)(  1.048,  2.0948864)(  1.054,  2.0935296)(  1.06,  2.092016)(
  1.066,  2.0903424)(  1.072,  2.088512)(  1.078,  2.0865216)(  1.084,
  2.0843776)(  1.09,  2.0820768)(  1.096,  2.0796192)(  1.102,  2.0770048)(
  1.108,  2.0742368)(  1.114,  2.0713152)(  1.12,  2.06824)(  1.126,  2.0650112
 )(  1.132,  2.0616288)(  1.138,  2.0580992)(  1.144,  2.054416)(  1.15,
  2.0505824)(  1.156,  2.0466016)(  1.162,  2.0424736)(  1.168,  2.0381984)(
  1.174,  2.033776)(  1.18,  2.0292096)(  1.186,  2.024496)(  1.192,  2.0196416
 )(  1.198,  2.0146464)(  1.204,  2.0095104)(  1.21,  2.0042336)(  1.216,
  1.9988192)(  1.222,  1.9932672)(  1.228,  1.9875808)(  1.234,  1.9817568)(
  1.24,  1.9758016)(  1.246,  1.9697152)(  1.252,  1.9634944)(  1.258,
  1.9571488)(  1.264,  1.9506752)(  1.27,  1.9440736)(  1.276,  1.9373472)(
  1.282,  1.9304992)(  1.288,  1.9235264)(  1.294,  1.9164352)(  1.3,
  1.9092288)(  1.306,  1.9019008)(  1.312,  1.8944608)(  1.318,  1.8869056)(
  1.324,  1.8792416)(  1.33,  1.8714656)(  1.336,  1.8635808)(  1.342,
  1.8555904)(  1.348,  1.8474944)(  1.354,  1.839296)(  1.36,  1.8309984)(
  1.366,  1.8226016)(  1.372,  1.8141088)(  1.378,  1.8055168)(  1.384,
  1.7968352)(  1.39,  1.7880608)(  1.396,  1.7792)(  1.402,  1.7702496)(  1.408
 ,  1.761216)(  1.414,  1.752096)(  1.42,  1.7428992)(  1.426,  1.7336224)(
  1.432,  1.7242688)(  1.438,  1.7148416)(  1.444,  1.7053408)(  1.45,
  1.6957696)(  1.456,  1.6861312)(  1.462,  1.6764256)(  1.468,  1.6666592)(
  1.474,  1.656832)(  1.48,  1.646944)(  1.486,  1.6369984)(  1.492,  1.6270016
 )(  1.498,  1.6169504)(  1.504,  1.606848)(  1.51,  1.5967008)(  1.516,
  1.5865088)(  1.522,  1.576272)(  1.528,  1.5659936)(  1.534,  1.55568)(  1.54
 ,  1.545328)(  1.546,  1.534944)(  1.552,  1.5245312)(  1.558,  1.5140864)(
  1.564,  1.503616)(  1.57,  1.4931232)(  1.576,  1.482608)(  1.582,  1.4720704
 )(  1.588,  1.46152)(  1.594,  1.4509536)(  1.6,  1.4403776)(  1.606,
  1.4297888)(  1.612,  1.4191936)(  1.618,  1.4085952)(  1.624,  1.3979936)(
  1.63,  1.3873888)(  1.636,  1.3767904)(  1.642,  1.3661952)(  1.648,
  1.3556064)(  1.654,  1.3450272)(  1.66,  1.3344576)(  1.666,  1.323904)(
  1.672,  1.3133664)(  1.678,  1.302848)(  1.684,  1.2923488)(  1.69,
  1.2818752)(  1.696,  1.271424)(  1.702,  1.2610016)(  1.708,  1.250608)(
  1.714,  1.2402464)(  1.72,  1.22992)(  1.726,  1.2196288)(  1.732,  1.209376)
 (  1.738,  1.1991648)(  1.744,  1.1889952)(  1.75,  1.1788704)(  1.756,
  1.1687936)(  1.762,  1.1587648)(  1.768,  1.148784)(  1.774,  1.1388608)(
  1.78,  1.1289888)(  1.786,  1.1191712)(  1.792,  1.1094144)(  1.798,
  1.0997184)(  1.804,  1.0900832)(  1.81,  1.080512)(  1.816,  1.071008)(
  1.822,  1.061568)(  1.828,  1.0521984)(  1.834,  1.042896)(  1.84,  1.0336704
 )(  1.846,  1.0245152)(  1.852,  1.0154336)(  1.858,  1.006432)(  1.864,
  0.9975072)(  1.87,  0.9886592)(  1.876,  0.9798944)(  1.882,  0.9712096)(
  1.888,  0.962608)(  1.894,  0.9540896)(  1.9,  0.9456576)(  1.906,  0.937312)
 (  1.912,  0.9290528)(  1.918,  0.9208832)(  1.924,  0.9128032)(  1.93,
  0.9048128)(  1.936,  0.896912)(  1.942,  0.889104)(  1.948,  0.8813888)(
  1.954,  0.8737664)(  1.96,  0.8662368)(  1.966,  0.8588032)(  1.972,
  0.8514656)(  1.978,  0.844224)(  1.984,  0.8370752)(  1.99,  0.8300256)(
  1.996,  0.8230688)(  2.002,  0.81621216)(  2.008,  0.80945152)(  2.014,
  0.80278848)(  2.02,  0.79622304)(  2.026,  0.78975488)(  2.032,  0.78338432)(
  2.038,  0.77711104)(  2.044,  0.77093568)(  2.05,  0.76485728)(  2.056,
  0.75887648)(  2.062,  0.75299232)(  2.068,  0.74720512)(  2.074,  0.74151424)
 (  2.08,  0.73591936)(  2.086,  0.73042048)(  2.092,  0.72501664)(  2.098,
  0.71970784)(  2.104,  0.71449312)(  2.11,  0.70937216)(  2.116,  0.70434464)(
  2.122,  0.6994096)(  2.128,  0.6945664)(  2.134,  0.6898144)(  2.14,
  0.68515296)(  2.146,  0.68058112)(  2.152,  0.67609856)(  2.158,  0.671704)(
  2.164,  0.6673968)(  2.17,  0.663176)(  2.176,  0.65904064)(  2.182,
  0.65499008)(  2.188,  0.65102304)(  2.194,  0.64713888)(  2.2,  0.64333664)(
  2.206,  0.63961504)(  2.212,  0.63597312)(  2.218,  0.63241024)(  2.224,
  0.6289248)(  2.23,  0.62551584)(  2.236,  0.62218272)(  2.242,  0.61892416)(
  2.248,  0.61573888)(  2.254,  0.6126256)(  2.26,  0.609584)(  2.266,
  0.60661216)(  2.272,  0.60370944)(  2.278,  0.60087456)(  2.284,  0.59810656)
 (  2.29,  0.59540384)(  2.296,  0.59276576)(  2.302,  0.59019136)(  2.308,
  0.58767872)(  2.314,  0.58522752)(  2.32,  0.58283616)(  2.326,  0.580504)(
  2.332,  0.57822944)(  2.338,  0.57601152)(  2.344,  0.57384928)(  2.35,
  0.57174176)(  2.356,  0.56968736)(  2.362,  0.56768576)(  2.368,  0.56573504)
 (  2.374,  0.56383488)(  2.38,  0.56198368)(  2.386,  0.5601808)(  2.392,
  0.55842496)(  2.398,  0.55671552)(  2.404,  0.5550512)(  2.41,  0.55343104)(
  2.416,  0.55185376)(  2.422,  0.55031904)(  2.428,  0.5488256)(  2.434,
  0.54737248)(  2.44,  0.54595872)(  2.446,  0.54458368)(  2.452,  0.54324608)(
  2.458,  0.54194528)(  2.464,  0.54068032)(  2.47,  0.53945024)(  2.476,
  0.53825472)(  2.482,  0.53709216)(  2.488,  0.53596256)(  2.494,  0.53486432)
 (  2.5,  0.53379744)(  2.506,  0.53276064)(  2.512,  0.531753216)(  2.518,
  0.53077456)(  2.524,  0.529823936)(  2.53,  0.528900576)(  2.536,
  0.528003808)(  2.542,  0.52713296)(  2.548,  0.526287328)(  2.554,
  0.525466304)(  2.56,  0.524669216)(  2.566,  0.523895456)(  2.572,
  0.523144416)(  2.578,  0.522415456)(  2.584,  0.521708032)(  2.59,
  0.521021536)(  2.596,  0.520355424)(  2.602,  0.51970912)(  2.608,
  0.519082144)(  2.614,  0.51847392)(  2.62,  0.517883936)(  2.626,
  0.517311712)(  2.632,  0.516756768)(  2.638,  0.516218592)(  2.644,
  0.515696768)(  2.65,  0.515190784)(  2.656,  0.514700224)(  2.662,
  0.514224672)(  2.668,  0.513763648)(  2.674,  0.5133168)(  2.68,  0.512883712
 )(  2.686,  0.512464)(  2.692,  0.512057216)(  2.698,  0.511663072)(  2.704,
  0.511281184)(  2.71,  0.510911136)(  2.716,  0.510552672)(  2.722,
  0.510205408)(  2.728,  0.509869024)(  2.734,  0.5095432)(  2.74,  0.509227616
 )(  2.746,  0.508922016)(  2.752,  0.508626048)(  2.758,  0.508339456)(  2.764
 ,  0.508061952)(  2.77,  0.50779328)(  2.776,  0.507533152)(  2.782,
  0.507281344)(  2.788,  0.5070376)(  2.794,  0.506801632)(  2.8,  0.506573248)
 (  2.806,  0.506352224)(  2.812,  0.506138336)(  2.818,  0.505931328)(  2.824,
  0.505731008)(  2.83,  0.505537216)(  2.836,  0.505349696)(  2.842,
  0.505168256)(  2.848,  0.504992768)(  2.854,  0.504822976)(  2.86,
  0.50465872)(  2.866,  0.504499872)(  2.872,  0.50434624)(  2.878,
  0.504197632)(  2.884,  0.50405392)(  2.89,  0.503914944)(  2.896,
  0.503780576)(  2.902,  0.503650624)(  2.908,  0.503524992)(  2.914,
  0.503403552)(  2.92,  0.503286112)(  2.926,  0.503172598)(  2.932,
  0.503062867)(  2.938,  0.5029568)(  2.944,  0.502854282)(  2.95,  0.502755194
 )(  2.956,  0.502659434)(  2.962,  0.502566886)(  2.968,  0.502477456)(  2.974
 ,  0.502391037)(  2.98,  0.502307536)(  2.986,  0.502226854)(  2.992,
  0.502148906)(  2.998,  0.5020736)(  3.004,  0.502000845)(  3.01,  0.501930566
 )(  3.016,  0.501862678)(  3.022,  0.501797104)(  3.028,  0.50173377)(  3.034,
  0.501672598)(  3.04,  0.50161352)(  3.046,  0.501556467)(  3.052,  0.50150137
 )(  3.058,  0.501448166)(  3.064,  0.501396794)(  3.07,  0.50134719)(  3.076,
  0.501299299)(  3.082,  0.501253059)(  3.088,  0.501208422)(  3.094,
  0.501165328)(  3.1,  0.501123728)(  3.106,  0.501083571)(  3.112,
  0.501044813)(  3.118,  0.501007402)(  3.124,  0.500971293)(  3.13,
  0.500936445)(  3.136,  0.500902813)(  3.142,  0.500870355)(  3.148,
  0.500839037)(  3.154,  0.500808816)(  3.16,  0.500779654)(  3.166,
  0.500751517)(  3.172,  0.500724368)(  3.178,  0.500698176)(  3.184,
  0.500672909)(  3.19,  0.500648531)(  3.196,  0.500625014)(  3.202,
  0.500602333)(  3.208,  0.500580451)(  3.214,  0.500559347)(  3.22,
  0.500538989)(  3.226,  0.500519357)(  3.232,  0.500500422)(  3.238,
  0.50048216)(  3.244,  0.50046455)(  3.25,  0.500447568)(  3.256,  0.500431194
 )(  3.262,  0.500415405)(  3.268,  0.500400179)(  3.274,  0.500385498)(  3.28,
  0.500371344)(  3.286,  0.500357699)(  3.292,  0.500344544)(  3.298,
  0.500331862)(  3.304,  0.500319638)(  3.31,  0.500307854)(  3.316,
  0.500296494)(  3.322,  0.500285545)(  3.328,  0.500274991)(  3.334,
  0.500264819)(  3.34,  0.500255016)(  3.346,  0.500245568)(  3.352,
  0.500236462)(  3.358,  0.500227688)(  3.364,  0.500219232)(  3.37,
  0.500211084)(  3.376,  0.500203232)(  3.382,  0.500195667)(  3.388,
  0.500188378)(  3.394,  0.500181355)(  3.4,  0.500174588)
 \end{pspicture}
\end{center}
\caption{\label{function} Dependence of the activation energy on the ratio
  between wall width and lattice parameter.}
\end{figure}

For intermediate cases we calculate $f(x)$ numerically. We calculate the
energy of a single wall in a 200 site system, both without
constraints ($\EM$) and with the constraint that $Q_{100}=0$ ($\ES$). 
The difference between the two energies divided by $\Delta E$
gives us $f(x)$, shown in Fig~\ref{function}. The figure shows that 
when the wall width is twice the lattice parameter the Peierls energy is
already practically zero.

In a crystal the width of a twin wall can be affected by two parameters: 
temperature and velocity. Elementary Landau-Ginzburg theory predicts that
the twin wall width should diverge as $T$ approaches $\TC$.\cite{Salje1990}
This prediction has been confirmed experimentally.\cite{Chrosch1999} 
As the temperature approaches $\TC$ the activation energy for wall motion
will decrease. The second factor which can affect the wall width is the 
velocity of the wall. If the speed of the wall $v$ approaches the velocity 
of sound in the material $c$ then the width of the wall is `Lorentz contracted'
by a factor of 
\begin{equation}
\sqrt{1-v^2/c^2}
\end{equation} 
As a wall accelerates the forces it experiences due to the lattice 
potential increase.

\section{Twin wall motion in CaTiO$_3$} 

To test the validity of the above approach we compared the value of the
activation energy calculated by the method described above with a direct
transition state energy calculation. We used an empirical potentials model to
investigate twin walls in CaTiO$_3$. We calculate the structure of the twin
walls of the system and calculate the Peierls energy and stress. 

\begin{figure}
 \begin{center}
\begin{pspicture}(5,3)
\rput[bl](0.2,0.2){\includegraphics[width=2cm]{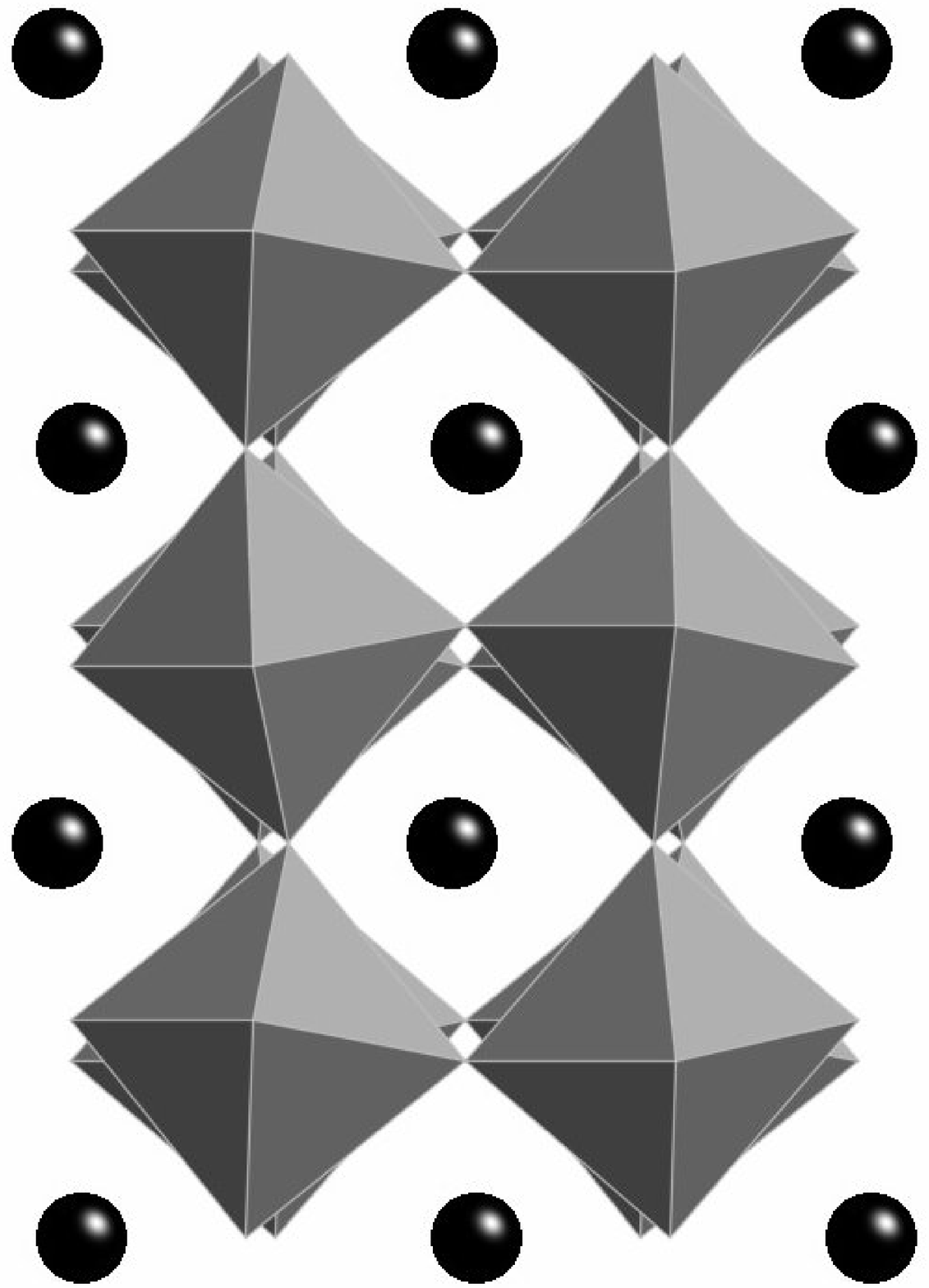}}
\rput[bl](3.0,0.6){\includegraphics[width=2cm]{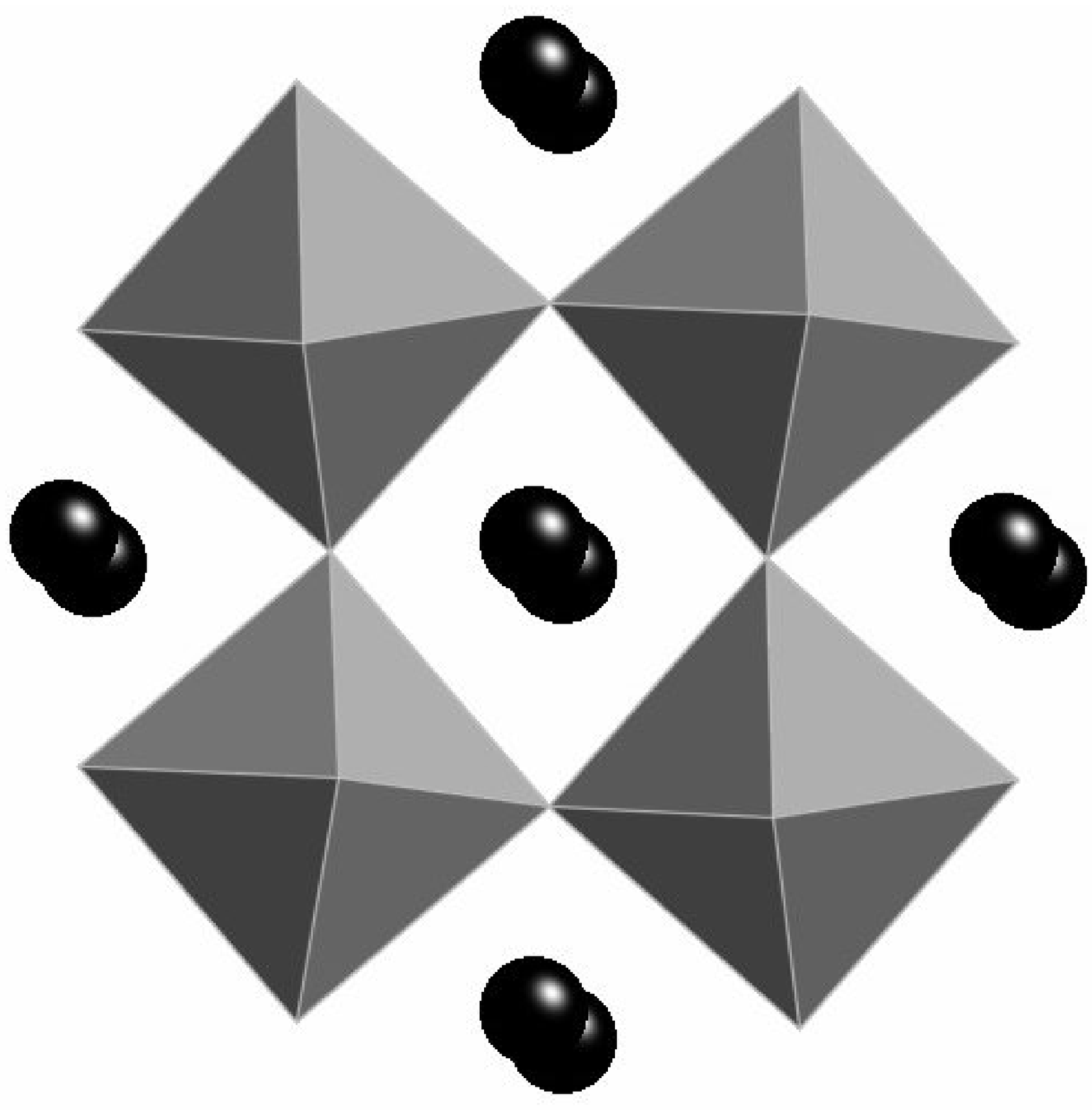}}
\psline{->}(0.1,0.1)(0.7,0.1)
\psline{->}(0.1,0.1)(0.1,0.7)
\uput{1pt}[ 0](0.7,0.1){\small $x$}
\uput{1pt}[90](0.1,0.7){\small $z$}
\psline{->}(2.9,0.5)(3.5,0.5)
\psline{->}(2.9,0.5)(2.9,1.1)
\uput{1pt}[ 0](3.5,0.5){\small $x$}
\uput{1pt}[90](2.9,1.1){\small $y$}
\end{pspicture}
 \end{center}
 \caption{\label{crystalstructure} Crystal structure of CaTiO$_3$, showing the
  coordinate system used in this work. Ca atoms are shown in black, the
  octahedra have O atoms at their vertices and Ti atoms at the centres.}
\end{figure}

CaTiO$_3$ is a ferroelastic, but not ferroelectric, perovskite. The crystal
structure consists of corner linked TiO$_6$ octahedra with Ca atoms
distributed between the octahedra. At high temperatures the crystal structure
is cubic but at room temperature the crystal structure is orthorhombic, with a
space group of $Pbnm$ and a Glazer octahedral tilt system of
$a^-a^-c^+$.\cite{Redfern1996,Glazer1972} The crystal structure and the
coordinate system used in this work is shown in Fig.~\ref{crystalstructure}.
When measurements of wall widths are given
below they are given in units of the pseudocubic unit cell, containing a
single formula unit.

\subsection{Structure of static twin walls.}

In this work we consider a ferroelastic wall perpendicular to the
$x$-axis. The structure of the wall can be described in terms of order
parameters and strains. The Glazer tilt system allows us to define order
parameters $(Q_x,Q_y,Q_z)$ associated with rotations of octahedra about the
$x$-, $y$-, and $z$-axes $(\theta_x,\theta_y,\theta_z)$. If the position
of an octahedron in the crystal is labelled by integers $(i_x,i_y,i_z)$, then
the order parameters are defined by  
\begin{align}
  \label{defineorder}
  \theta_x &=Q_x (-1)^{i_x+i_y+i_z}\\
  \theta_y &=Q_y (-1)^{i_x+i_y+i_z}\\
  \theta_z &=Q_z (-1)^{i_x+i_y}
\end{align}
The compatibility conditions limit the strains which can vary across an
interface. For an interface perpendicular to the $x$-axis only the strains
$\epsilon_{xx}$, $\epsilon_{xy}$ and $\epsilon_{zx}$ can be non-zero. 
Furthermore the symmetry of the crystal 
constrains the strain $\epsilon_{zx}$ to be zero. The strain $\epsilon_{xy}$ is the ferroelastic
strain. This changes sign across a ferroelastic wall. The strain
$\epsilon_{xx}$ is a secondary strain, which only takes non-zero values within the
wall.  

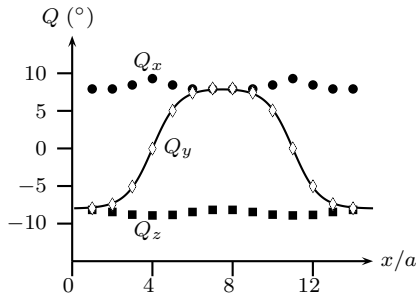
\begin{figure}
\begin{center}
\begin{pspicture}(          6.000000,          4.000000)
\psline[linecolor=black]{->}(          1.000000,          0.500000)(          5.000000,          0.500000)
\psline[linecolor=black]{->}(          1.000000,          0.500000)(          1.000000,          3.500000)
\uput{3pt}[0](          5.000000,          0.500000){\footnotesize $x/a$                                                       }
\uput{3pt}[90](          1.000000,          3.500000){\footnotesize$Q\; (^\circ)$                                              }
\psline[linecolor=black]{-}(          1.000000,          0.500000)(          1.000000,          0.300000)
\uput{1pt}[270](          1.000000,          0.300000){\footnotesize$0$       }
\psline[linecolor=black]{-}(          2.066667,          0.500000)(          2.066667,          0.300000)
\uput{1pt}[270](          2.066667,          0.300000){\footnotesize$4$       }
\psline[linecolor=black]{-}(          3.133333,          0.500000)(          3.133333,          0.300000)
\uput{1pt}[270](          3.133333,          0.300000){\footnotesize$8$       }
\psline[linecolor=black]{-}(          4.200000,          0.500000)(          4.200000,          0.300000)
\uput{1pt}[270](          4.200000,          0.300000){\footnotesize$12$      }
\psline[linecolor=black](          1.000000,          1.000000)(          0.800000,          1.000000)
\uput{1pt}[180](          0.800000,          1.000000){\footnotesize$-10$     }
\psline[linecolor=black](          1.000000,          1.500000)(          0.800000,          1.500000)
\uput{1pt}[180](          0.800000,          1.500000){\footnotesize$-5$      }
\psline[linecolor=black](          1.000000,          2.000000)(          0.800000,          2.000000)
\uput{1pt}[180](          0.800000,          2.000000){\footnotesize$0$       }
\psline[linecolor=black](          1.000000,          2.500000)(          0.800000,          2.500000)
\uput{1pt}[180](          0.800000,          2.500000){\footnotesize$5$       }
\psline[linecolor=black](          1.000000,          3.000000)(          0.800000,          3.000000)
\uput{1pt}[180](          0.800000,          3.000000){\footnotesize$10$      }
\psdots[linecolor=black,dotstyle=*,        dotscale=1.0]%
(          1.266667,          2.792578)%
(          1.533333,          2.792126)%
(          1.800000,          2.844162)%
(          2.066667,          2.928707)%
(          2.333333,          2.844815)%
(          2.600000,          2.792212)%
(          2.866667,          2.792561)%
(          3.133333,          2.792578)%
(          3.400000,          2.792131)%
(          3.666667,          2.844167)%
(          3.933333,          2.928707)%
(          4.200000,          2.844820)%
(          4.466667,          2.792217)%
(          4.733333,          2.792561)%

\psdots[linecolor=black,dotstyle=square*,  dotscale=1.0]%
(          1.266667,          1.184234)%
(          1.533333,          1.153650)%
(          1.800000,          1.117616)%
(          2.066667,          1.108764)%
(          2.333333,          1.117462)%
(          2.600000,          1.153380)%
(          2.866667,          1.184125)%
(          3.133333,          1.184234)%
(          3.400000,          1.153650)%
(          3.666667,          1.117622)%
(          3.933333,          1.108764)%
(          4.200000,          1.117467)%
(          4.466667,          1.153380)%
(          4.733333,          1.184131)%

\pscurve[linecolor=black]%
(          1.026667,          1.203956)%
(          1.053333,          1.204612)%
(          1.080000,          1.205377)%
(          1.106667,          1.206267)%
(          1.133333,          1.207305)%
(          1.160000,          1.208513)%
(          1.186667,          1.209920)%
(          1.213333,          1.211558)%
(          1.240000,          1.213464)%
(          1.266667,          1.215682)%
(          1.293333,          1.218260)%
(          1.320000,          1.221257)%
(          1.346667,          1.224737)%
(          1.373333,          1.228777)%
(          1.400000,          1.233463)%
(          1.426667,          1.238893)%
(          1.453333,          1.245179)%
(          1.480000,          1.252446)%
(          1.506667,          1.260837)%
(          1.533333,          1.270509)%
(          1.560000,          1.281637)%
(          1.586667,          1.294413)%
(          1.613333,          1.309045)%
(          1.640000,          1.325755)%
(          1.666667,          1.344777)%
(          1.693333,          1.366352)%
(          1.720000,          1.390720)%
(          1.746667,          1.418115)%
(          1.773333,          1.448750)%
(          1.800000,          1.482807)%
(          1.826667,          1.520422)%
(          1.853333,          1.561665)%
(          1.880000,          1.606531)%
(          1.906667,          1.654917)%
(          1.933333,          1.706617)%
(          1.960000,          1.761313)%
(          1.986667,          1.818572)%
(          2.013333,          1.877860)%
(          2.040000,          1.938554)%
(          2.066667,          1.999966)%
(          2.093333,          2.061378)%
(          2.120000,          2.122069)%
(          2.146667,          2.181353)%
(          2.173333,          2.238607)%
(          2.200000,          2.293294)%
(          2.226667,          2.344985)%
(          2.253333,          2.393359)%
(          2.280000,          2.438210)%
(          2.306667,          2.479436)%
(          2.333333,          2.517029)%
(          2.360000,          2.551061)%
(          2.386667,          2.581667)%
(          2.413333,          2.609027)%
(          2.440000,          2.633354)%
(          2.466667,          2.654881)%
(          2.493333,          2.673847)%
(          2.520000,          2.690492)%
(          2.546667,          2.705048)%
(          2.573333,          2.717735)%
(          2.600000,          2.728759)%
(          2.626667,          2.738310)%
(          2.653333,          2.746560)%
(          2.680000,          2.753663)%
(          2.706667,          2.759756)%
(          2.733333,          2.764962)%
(          2.760000,          2.769387)%
(          2.786667,          2.773122)%
(          2.813333,          2.776247)%
(          2.840000,          2.778829)%
(          2.866667,          2.780925)%
(          2.893333,          2.782579)%
(          2.920000,          2.783829)%
(          2.946667,          2.784703)%
(          2.973333,          2.785219)%
(          3.000000,          2.785390)%
(          3.026667,          2.785219)%
(          3.053334,          2.784703)%
(          3.080000,          2.783829)%
(          3.106667,          2.782579)%
(          3.133333,          2.780925)%
(          3.160000,          2.778829)%
(          3.186667,          2.776247)%
(          3.213333,          2.773122)%
(          3.240000,          2.769387)%
(          3.266667,          2.764962)%
(          3.293334,          2.759756)%
(          3.320000,          2.753663)%
(          3.346667,          2.746560)%
(          3.373333,          2.738310)%
(          3.400000,          2.728759)%
(          3.426667,          2.717735)%
(          3.453333,          2.705048)%
(          3.480000,          2.690492)%
(          3.506667,          2.673847)%
(          3.533333,          2.654881)%
(          3.560000,          2.633354)%
(          3.586667,          2.609027)%
(          3.613333,          2.581666)%
(          3.640000,          2.551061)%
(          3.666667,          2.517029)%
(          3.693333,          2.479435)%
(          3.720000,          2.438210)%
(          3.746667,          2.393359)%
(          3.773334,          2.344985)%
(          3.800000,          2.293294)%
(          3.826667,          2.238606)%
(          3.853333,          2.181353)%
(          3.880000,          2.122069)%
(          3.906667,          2.061378)%
(          3.933333,          1.999966)%
(          3.960000,          1.938554)%
(          3.986667,          1.877860)%
(          4.013333,          1.818572)%
(          4.040000,          1.761312)%
(          4.066667,          1.706617)%
(          4.093333,          1.654917)%
(          4.120000,          1.606531)%
(          4.146667,          1.561665)%
(          4.173334,          1.520421)%
(          4.200000,          1.482807)%
(          4.226667,          1.448750)%
(          4.253333,          1.418115)%
(          4.280000,          1.390720)%
(          4.306667,          1.366352)%
(          4.333333,          1.344777)%
(          4.360000,          1.325755)%
(          4.386667,          1.309045)%
(          4.413333,          1.294413)%
(          4.440000,          1.281637)%
(          4.466667,          1.270509)%
(          4.493333,          1.260837)%
(          4.520000,          1.252446)%
(          4.546667,          1.245179)%
(          4.573333,          1.238893)%
(          4.600000,          1.233463)%
(          4.626667,          1.228777)%
(          4.653333,          1.224737)%
(          4.680000,          1.221257)%
(          4.706667,          1.218260)%
(          4.733333,          1.215682)%
(          4.760000,          1.213464)%
(          4.786666,          1.211558)%
(          4.813334,          1.209920)%
(          4.840000,          1.208513)%
(          4.866667,          1.207305)%
(          4.893333,          1.206267)%
(          4.920000,          1.205377)%
(          4.946667,          1.204612)%
(          4.973333,          1.203956)%
(          5.000000,          1.203393)%

\psdots[linecolor=black,dotstyle=diamond,  dotscale=1.0]%
(          1.266667,          1.202437)%
(          1.533333,          1.258100)%
(          1.800000,          1.494376)%
(          2.066667,          1.998018)%
(          2.333333,          2.503011)%
(          2.600000,          2.741121)%
(          2.866667,          2.797448)%
(          3.133333,          2.797563)%
(          3.400000,          2.741900)%
(          3.666667,          2.505618)%
(          3.933333,          2.001982)%
(          4.200000,          1.496989)%
(          4.466667,          1.258879)%
(          4.733333,          1.202546)%

\rput(          2.066667,          3.150000){\footnotesize$Q_x$                                                       }
\rput(          2.066667,          0.900000){\footnotesize$Q_z$                                                       }
\rput(          2.466667,          2.000000){\footnotesize$Q_y$                                                       }
\end{pspicture}
\end{center}
\caption{\label{wallorder} The behaviour of the order parameters across the
  wall. The order parameters are described in
  equation~\ref{defineorder}. $Q_y$ shows a hyperbolic tangent variation
  across the wall, and $Q_x$ and $Q_y$ show anomalies at the wall. }
\end{figure}

\begin{figure}
\begin{center}
\begin{pspicture}(          6.000000,          4.000000)
\psline[linecolor=black]{->}(          1.000000,          0.500000)(          5.000000,          0.500000)
\psline[linecolor=black]{->}(          1.000000,          0.500000)(          1.000000,          3.500000)
\uput{3pt}[0](          5.000000,          0.500000){\footnotesize $x/a$                                                       }
\uput{3pt}[90](          1.000000,          3.500000){\footnotesize$\epsilon \; (10^{-3})$                                     }
\psline[linecolor=black]{-}(          1.000000,          0.500000)(          1.000000,          0.300000)
\uput{1pt}[270](          1.000000,          0.300000){\footnotesize$0$       }
\psline[linecolor=black]{-}(          2.066667,          0.500000)(          2.066667,          0.300000)
\uput{1pt}[270](          2.066667,          0.300000){\footnotesize$4$       }
\psline[linecolor=black]{-}(          3.133333,          0.500000)(          3.133333,          0.300000)
\uput{1pt}[270](          3.133333,          0.300000){\footnotesize$8$       }
\psline[linecolor=black]{-}(          4.200000,          0.500000)(          4.200000,          0.300000)
\uput{1pt}[270](          4.200000,          0.300000){\footnotesize$12$      }
\psline[linecolor=black](          1.000000,          0.800000)(          0.800000,          0.800000)
\uput{1pt}[180](          0.800000,          0.800000){\footnotesize$-4$      }
\psline[linecolor=black](          1.000000,          1.400000)(          0.800000,          1.400000)
\uput{1pt}[180](          0.800000,          1.400000){\footnotesize$-2$      }
\psline[linecolor=black](          1.000000,          2.000000)(          0.800000,          2.000000)
\uput{1pt}[180](          0.800000,          2.000000){\footnotesize$0$       }
\psline[linecolor=black](          1.000000,          2.600000)(          0.800000,          2.600000)
\uput{1pt}[180](          0.800000,          2.600000){\footnotesize$2$       }
\psline[linecolor=black](          1.000000,          3.200000)(          0.800000,          3.200000)
\uput{1pt}[180](          0.800000,          3.200000){\footnotesize$4$       }
\psdots[linecolor=black,dotstyle=*,        dotscale=1.0]%
(          1.400000,          1.717700)%
(          1.666667,          1.776500)%
(          1.933333,          2.729900)%
(          2.200000,          2.427500)%
(          2.466667,          1.966100)%
(          2.733333,          1.678700)%
(          3.000000,          1.703600)%
(          3.266667,          1.717700)%
(          3.533333,          1.776500)%
(          3.800000,          2.729900)%
(          4.066667,          2.427500)%
(          4.333333,          1.965800)%
(          4.600000,          1.678700)%

\psdots[linecolor=black,dotstyle=diamond,  dotscale=1.0]%
(          1.400000,          0.756800)%
(          1.666667,          1.246700)%
(          1.933333,          1.897400)%
(          2.200000,          2.037500)%
(          2.466667,          2.874800)%
(          2.733333,          3.189500)%
(          3.000000,          3.282500)%
(          3.266667,          3.243500)%
(          3.533333,          2.753300)%
(          3.800000,          2.102600)%
(          4.066667,          1.962200)%
(          4.333333,          1.124900)%
(          4.600000,          0.810200)%

\rput(          1.666667,          2.750000){\footnotesize$\epsilon_{xx}$                                             }
\rput(          3.133333,          2.900000){\footnotesize$\epsilon_{xy}$                                             }
\end{pspicture}
\end{center}
\caption{\label{wallstrains} Strain behaviour across the twin wall. The
  ferroelastic 
  strain $\epsilon_{xy}$ follows the order parameter $Q_y$, while the
  secondary  
  strain $\epsilon_{xx}$ shows an anomaly within the wall.}
\end{figure}
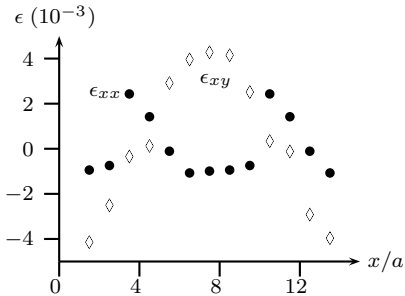

Calleja et al.\cite{Calleja2003} developed an empirical potential set for
this mineral and 
investigated the interaction between oxygen vacancies and twin walls in a
configuration containing $26 \times 10 \times 6$ 
cells. The authors simulated a single domain structure and then rotated part
of their configuration through $90^{\circ}$ to generate twin walls. This
procedure generates an interface consisting of the combination of a
ferroelastic twin wall (with an order parameter $Q_y$) with
an antiphase boundary (with an order parameter $Q_z$). 
These two types of wall can exist independently so in this work we
consider simple 
ferroelastic twin walls in a system of $14 \times 6 \times 4$ octahedra
implemented in \code{DL\_POLY}~\cite{dlpoly} using Calleja et al.'s potential
set. Periodic boundary conditions make it impossible to simulate a single twin
wall so instead we simulate a system with two walls. The order parameters 
and strains in the walls relaxed at absolute zero are shown in 
Figs.~\ref{wallorder} and~\ref{wallstrains}. The order parameter $Q_y$ and
the shear strain $\epsilon_{xy}$ change sign across the walls. Fitting $Q_y$ to
a hyperbolic tangent profile gives a wall width $w=1.3a$. $Q_x$, $Q_z$, and
$\epsilon_{xx}$ show anomalies across the wall. This is the behaviour expected
from secondary order parameters.\cite{Lee2003a}

\subsection{Activation energy for twin wall motion in CaTiO$_3$.}

We compare two methods of calculating the Peierls energy $\EP$. The first
method is 
an indirect calculation using equation~\ref{generalequation}. Secondly we
perform a direct calculation of the transition state energy. These two
energies are in good agreement. We also calculate the Peierls stress $\SP$
which is the maximum restoring stress the wall can exert.   

To calculate the Peierls energy $\EP$ using
equation~\ref{generalequation}, we first calculate $\Delta E$ from the
interfacial energy $\gamma$ of the twin walls. From $\Delta E$ and the value
of the function $f$ for $w/a=1.3$ we can calculate $\EP$. Finally, assuming a
sinusoidal variation of the energy as the wall moves 
through the lattice we can calculate the Peierls stress $\SP$. The stages of
the calculation are summarised in Tab.~\ref{peierls} 

\begin{table}[!h]
\begin{tabular}{ l  r@{.}l }
\hline
\multicolumn{3}{c}{Indirect Calculation}\\
\hline
$\gamma$   &   0&$116\unit{J}\unit{m}^{-2}$ \\
$\Delta E$ &   0&$034\unit{J}\unit{m}^{-2}$ \\
$w/a$      &   1&$13$ \\
$f(w/a)$   &   0&$016$\\
$\EP$      &   0&$530 \unit{mJ}\unit{m}^{-2}$\\
$\SP$      &   4&$35\unit{MPa}$\\
\hline
\multicolumn{3}{c}{Direct Calculation}\\
\hline
$\EP$      &   0&$313 \unit{mJ}\unit{m}^{-2}$\\
$\SP$      &   2&$57\unit{MPa}$\\
\hline
\end{tabular}
\caption{\label{peierls} Indirect and direct calculation of the
  activation energy and Peierls stress. The results of the 
  calculations are in good agreement.} 
\end{table}

To calculate the Peierls energy directly we perform a transition
state calculation on our system. We start with initial and final states which
differ in that the two twin walls of the system are each translated by one lattice
parameter in the same direction. The difference in atomic coordinates between 
these two states
defines the reaction coordinate. We moved the system from the initial to the
final state in 100 steps, relaxing all the degrees of freedom perpendicular to
the reaction coordinate. At each step we calculated the force along the reaction
coordinate and, by numerical integration of the work done by that force, the
Peierls energy and stress. (This approach was necessary because \code{DL\_POLY}
cannot directly resolve the energy differences involved.) Again the results of
these calculations are summarised in Tab.~\ref{peierls}.  

The agreement between our two results is very good---less than a factor of
two---especially given the small value of the Peierls energy compared with the
interaction energy of a twin wall with an oxygen vacancy, which, as noted
above, is of the order of $1\unit{eV}$. The residual 
discrepancy may be due to the complexity of the
system. Equation~\ref{generalequation} was developed for a domain wall which
can be described by a single order parameter. As shown in
Figs.~\ref{wallorder} and~\ref{wallstrains} at least five parameters show
anomalies across the wall. The energies of these anomalies may lead to an
overestimation of $\Delta E$ calculated from the interfacial energy $\gamma$
of the wall.  

\subsection{Simulation of a moving domain wall}

Our results suggest that if a pressure greater than $\SP \approx 3\unit{MPa}$
is applied to a twin wall it will move freely, rather than as a thermally
activated process. In this section we demonstrate that this is the case by
molecular dynamics simulation. Working in an $NVT$ ensemble we shear the
system to generate a force on the walls and observe their motion.

In order to calculate the force on the wall generated by a shear stress we
need to calculate the Eshelby force on the wall.\cite{Eshelby1975} The stress
on a wall $\SW$ generated by an externally applied shear stress $\sigma_{xy}$
is given by
\begin{equation}
\SW=4\sigma_{xy}\epsilon_{xy}
\end{equation}
where $\epsilon_{xy}$ is the spontaneous strain of the transition. For
CaTiO$_3$ $\epsilon_{xy}=4\times 10^{-3}$ (see Fig.~\ref{wallstrains}). 

\begin{table}
\begin{center}
\begin{tabular}{l r@{.}l}
\hline
Timestep                         &       1&$0\unit{fs}$\\
Thermostat relaxation time       &       0&$5\unit{ps}$\\
Simulation duration              &      10&$0\unit{ps}$\\
Initial shear stress on crystal  &       6&$0\unit{GPa}$\\
Initial pressure on walls        &     100&$0\unit{MPa}$\\
Peierls stress                   &       3&$0\unit{MPa}$\\
\hline
\end{tabular} 
\end{center}
\caption{\label{movingwall}Parameters used in the simulation of a moving twin
  wall. The stresses acting on the system and the wall are only initial
  values. As the walls move in response to the forces these stresses will
  relax. }
\end{table}

\begin{figure}
\begin{center}
\begin{pspicture}(          6.000000,          4.000000)
\psline[linecolor=black]{->}(          1.000000,          0.500000)(          5.000000,          0.500000)
\psline[linecolor=black]{->}(          1.000000,          0.500000)(          1.000000,          3.500000)
\uput{3pt}[0](          5.000000,          0.500000){\footnotesize $t\unit{(ps)}$                                              }
\uput{3pt}[90](          1.000000,          3.500000){\footnotesize$x/a$                                                       }
\psline[linecolor=black]{-}(          1.000000,          0.500000)(          1.000000,          0.300000)
\uput{1pt}[270](          1.000000,          0.300000){\footnotesize$0$       }
\psline[linecolor=black]{-}(          1.727273,          0.500000)(          1.727273,          0.300000)
\uput{1pt}[270](          1.727273,          0.300000){\footnotesize$2$       }
\psline[linecolor=black]{-}(          2.454545,          0.500000)(          2.454545,          0.300000)
\uput{1pt}[270](          2.454545,          0.300000){\footnotesize$4$       }
\psline[linecolor=black]{-}(          3.181818,          0.500000)(          3.181818,          0.300000)
\uput{1pt}[270](          3.181818,          0.300000){\footnotesize$6$       }
\psline[linecolor=black]{-}(          3.909091,          0.500000)(          3.909091,          0.300000)
\uput{1pt}[270](          3.909091,          0.300000){\footnotesize$8$       }
\psline[linecolor=black]{-}(          4.636364,          0.500000)(          4.636364,          0.300000)
\uput{1pt}[270](          4.636364,          0.300000){\footnotesize$10$      }
\psline[linecolor=black](          1.000000,          0.875000)(          0.800000,          0.875000)
\uput{1pt}[180](          0.800000,          0.875000){\footnotesize$-6$      }
\psline[linecolor=black](          1.000000,          1.437500)(          0.800000,          1.437500)
\uput{1pt}[180](          0.800000,          1.437500){\footnotesize$-3$      }
\psline[linecolor=black](          1.000000,          2.000000)(          0.800000,          2.000000)
\uput{1pt}[180](          0.800000,          2.000000){\footnotesize$0$       }
\psline[linecolor=black](          1.000000,          2.562500)(          0.800000,          2.562500)
\uput{1pt}[180](          0.800000,          2.562500){\footnotesize$3$       }
\psline[linecolor=black](          1.000000,          3.125000)(          0.800000,          3.125000)
\uput{1pt}[180](          0.800000,          3.125000){\footnotesize$6$       }
\pscurve[linecolor=black]%
(          1.072727,          0.845733)%
(          1.145455,          0.846546)%
(          1.218182,          0.848513)%
(          1.290909,          0.853252)%
(          1.363636,          0.860678)%
(          1.436364,          0.870136)%
(          1.509091,          0.880804)%
(          1.581818,          0.891570)%
(          1.654545,          0.903832)%
(          1.727273,          0.918064)%
(          1.800000,          0.934862)%
(          1.872727,          0.952102)%
(          1.945455,          0.967443)%
(          2.018182,          0.982893)%
(          2.090909,          1.000480)%
(          2.163636,          1.019494)%
(          2.236364,          1.038103)%
(          2.309091,          1.055629)%
(          2.381818,          1.073662)%
(          2.454545,          1.091815)%
(          2.527273,          1.110504)%
(          2.600000,          1.130234)%
(          2.672727,          1.150779)%
(          2.745455,          1.168728)%
(          2.818182,          1.184809)%
(          2.890909,          1.202292)%
(          2.963636,          1.220110)%
(          3.036364,          1.237625)%
(          3.109091,          1.255335)%
(          3.181818,          1.273624)%
(          3.254545,          1.290894)%
(          3.327273,          1.306242)%
(          3.400000,          1.321562)%
(          3.472727,          1.336351)%
(          3.545455,          1.350695)%
(          3.618182,          1.364519)%
(          3.690909,          1.378392)%
(          3.763636,          1.392809)%
(          3.836364,          1.407549)%
(          3.909091,          1.420843)%
(          3.981818,          1.435160)%
(          4.054546,          1.449366)%
(          4.127273,          1.460935)%
(          4.200000,          1.471730)%
(          4.272727,          1.482539)%
(          4.345455,          1.491473)%
(          4.418182,          1.498982)%
(          4.490909,          1.506282)%
(          4.563636,          1.513223)%
(          4.636364,          1.518725)%

\pscurve[linecolor=black]%
(          1.072727,          3.112868)%
(          1.145455,          3.111245)%
(          1.218182,          3.107236)%
(          1.290909,          3.100309)%
(          1.363636,          3.091379)%
(          1.436364,          3.082763)%
(          1.509091,          3.074704)%
(          1.581818,          3.064804)%
(          1.654545,          3.053318)%
(          1.727273,          3.039083)%
(          1.800000,          3.024552)%
(          1.872727,          3.009881)%
(          1.945455,          2.993403)%
(          2.018182,          2.978044)%
(          2.090909,          2.962519)%
(          2.163636,          2.943770)%
(          2.236364,          2.925522)%
(          2.309091,          2.909064)%
(          2.381818,          2.893303)%
(          2.454545,          2.877757)%
(          2.527273,          2.860240)%
(          2.600000,          2.843485)%
(          2.672727,          2.828611)%
(          2.745455,          2.813030)%
(          2.818182,          2.797965)%
(          2.890909,          2.784709)%
(          2.963636,          2.769447)%
(          3.036364,          2.754065)%
(          3.109091,          2.739516)%
(          3.181818,          2.727630)%
(          3.254545,          2.716691)%
(          3.327273,          2.704684)%
(          3.400000,          2.692797)%
(          3.472727,          2.683130)%
(          3.545455,          2.674327)%
(          3.618182,          2.666055)%
(          3.690909,          2.658178)%
(          3.763636,          2.649917)%
(          3.836364,          2.641060)%
(          3.909091,          2.632826)%
(          3.981818,          2.625588)%
(          4.054546,          2.619070)%
(          4.127273,          2.611696)%
(          4.200000,          2.604261)%
(          4.272727,          2.598640)%
(          4.345455,          2.593219)%
(          4.418182,          2.587149)%
(          4.490909,          2.580673)%
(          4.563636,          2.574277)%
(          4.636364,          2.569052)%

\end{pspicture}
\end{center}
\caption{\label{wallmotion}Observed motion of the two twin walls of the 
  simulated system.}
\end{figure}
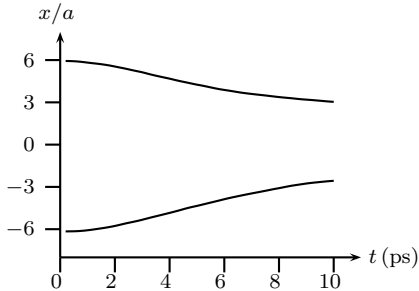

We started with a configuration of $26\times 10 \times 6$ octahedra from the
simulation of Calleja et al,
containing, as noted above, both ferroelastic walls and 
antiphase boundaries. On annealing at $10\unit{K}$ using \code{DL\_POLY} the
antiphase 
boundaries spontaneously moved together and annihilated each other, leaving
only the ferroelastic walls. \code{DL\_POLY} does not allow the direct imposition of
a constant shear stress so instead we sheared the whole system (both
coordinates and velocities) through an
angle of $0.3^{\circ}$, generating an initial shear stress. (The $NVT$ ensemble
prevents the relaxation of this stress by a macroscopic shear of the system.)
The Eshelby force 
on the wall exceeded the Peierls stress and so motion of the wall was
observed. The parameters of the simulation are summarised in
Tab.~\ref{movingwall}. In response to these forces the walls move as shown in
Fig.~\ref{wallmotion}. Initially the walls accelerate because the pressure
acting on them is higher than the Peierls stress. The walls traverse several
unit cells and then decelerate, as the stress acting on them decreases.

\section{Conclusions}

A complete picture of twin wall motion in ferroelastic and ferroelectric
materials would shed light on questions such as the fatigue problem in
ferroelectric memories and the contribution of twin wall motion to the
seismic properties of the Earth's lower mantle. Such a picture requires an
understanding of the energy landscape through which the twin wall moves in the
presence and absence of point defects. We have shown that the Peierls energy 
and stress of a ferroic material can be 
accurately estimated using an indirect approach by mapping the system on to a 
one order parameter model.\\

\begin{acknowledgements}
The authors wish to thank Mark Calleja for providing the atomic configuration
used as a starting point in section~III~C. 
\end{acknowledgements}

\end{document}